\documentclass[journal]{IEEEtran}
\ifCLASSINFOpdf
\else
 \usepackage[dvips]{graphicx}
\fi
\usepackage{url}
\usepackage{graphicx}
\usepackage{cite}
\usepackage{amsmath,amssymb,amsfonts}
\usepackage{algorithmic}
\usepackage{graphicx}
\usepackage{textcomp}
\usepackage{xcolor}
\usepackage{bm}
\usepackage{color}
\usepackage{subfigure}
\usepackage{makecell}
\usepackage{algorithm}
\usepackage{microtype}

\begin{document}
\title{Flexible Intelligent Metasurfaces for Enhancing\\ Secure MISO Communications}
\author{Zihao Teng, \IEEEmembership{Graduate Student Member, IEEE}, Zihao Zhao, Jiancheng An, \IEEEmembership{Senior Member, IEEE},\\ Lu Gan, \IEEEmembership{Member, IEEE}, Arumugam Nallanathan, \IEEEmembership{Fellow, IEEE}, and Bo Ai, \IEEEmembership{Fellow, IEEE}\vspace{-5mm}
\thanks{Z. Teng, Z. Zhao and L. Gan are with the School of Information and Communication Engineering, University of Electronic Science and Technology of China (UESTC), Chengdu, 611731, China. L. Gan is also with the Yibin Institute of UESTC, Yibin, Sichuan, 643000, China (e-mail: tzhuestc@163.com; 202111012024@std.uestc.edu.cn; ganlu@uestc.edu.cn).}
\thanks{J. An is with the School of Electronic Science and Engineering, UESTC, Chengdu, 611731, China (e-mail:  jiancheng.an@uestc.edu.cn).}
\thanks{A. Nallanathan is with the School of Electronic Engineering and Computer Science, Queen Mary University of London, London (e-mail: a.nallanathan@qmul.ac.uk).}
\thanks{B. Ai is with the School of Electronic and Information Engineering, Beijing
Jiaotong University, Beijing 100044, China (e-mail: boai@bjtu.edu.cn).}
}
\markboth{DRAFT}{DRAFT}
\maketitle
\begin{abstract}
This paper investigates flexible intelligent metasurface (FIM)-aided secure multiple-input single-output (MISO) communications in the presence of a multi-antenna eavesdropper. By exploiting the morphing capability of the FIM, the transmit array geometry can be dynamically reconfigured to reshape the legitimate and eavesdropping channels. We formulate a secrecy rate maximization (SRM) problem by jointly optimizing the transmit covariance matrix and the FIM surface shape. For a given FIM surface shape, the optimal transmit covariance matrix is shown to be rank one and is obtained in closed form from the dominant generalized eigenvector. For a fixed transmit covariance matrix, a projected gradient ascent (PGA) method is developed to optimize the FIM shape. Based on these two subproblems, a block coordinate ascent (BCA) algorithm is proposed. Moreover, for the special line-of-sight (LoS) scenario, we derive a globally optimal closed-form joint transmit covariance matrix and FIM surface shape design under a sufficient morphing-range condition. Numerical results demonstrate that FIM effectively improves the secrecy rate over a conventional rigid antenna array (RAA). Moreover, FIM achieves a transmit-power saving of about $2.5$ dB compared with the RAA benchmark at the same secrecy rate.
\end{abstract}
\begin{IEEEkeywords}
	Flexible intelligent metasurface (FIM), physical layer secrecy (PLS), multiple-input single-output (MISO), surface shape morphing, beamforming.
\end{IEEEkeywords}
\IEEEpeerreviewmaketitle
\section{Introduction} 
\IEEEPARstart{T}{he} broadcast nature of wireless propagation makes confidential information inherently vulnerable to eavesdropping \cite{niu2026survey}. To address this issue, reconfigurable intelligent surfaces (RISs) have emerged as a promising technology for enhancing physical layer security (PLS) by reshaping wireless propagation through programmable metasurface elements. By jointly optimizing the transmit beamforming and RIS reflection coefficients, the legitimate signal can be enhanced while the information leakage toward eavesdroppers is suppressed \cite{zhang2025rissecure}. To further enhance wave-domain processing capability, stacked intelligent metasurfaces (SIMs) have recently been developed by cascading multiple programmable metasurface layers \cite{an2023stacked}. Compared with single-layer RISs, SIMs can perform more sophisticated electromagnetic transformations through multilayer signal propagation and have shown potential for secure transmission design. For example, the authors in \cite{niu2025efficient} investigated a secure single-input single-output (SISO) transmission system, where an SIM was deployed at the transmitter to jointly realize modulation, beamforming, and artificial-noise generation in the electromagnetic wave domain. The work in \cite{kavianinia2025secrecy} extended SIM-assisted secure transmission to multiuser multiple-input single-output (MISO) systems, where the base-station beamforming vectors and the discrete phase shifts of multiple metasurface layers were jointly optimized for sum secrecy rate maximization (SRM). However, both RISs and SIMs mainly manipulate the electromagnetic responses of fixed meta-atoms, which may limit their ability to reduce the spatial correlation between the legitimate and eavesdropping channels, especially when the eavesdropper is equipped with multiple antennas. 

Driven by recent advances in flexible materials, flexible intelligent metasurfaces (FIMs) have emerged as a new architecture capable of dynamically reconfiguring their three-dimensional surface shapes \cite{an2025emer}. Unlike RISs and SIMs with fixed geometries \cite{zhang2025rissecure,kavianinia2025secrecy}, an FIM reshapes wireless channels through morphing-induced, direction-dependent phase responses \cite{anfim2025}. Recent studies have demonstrated the benefits of FIM morphing for enhancing communication \cite{anfim2025,kumar2026fimstatcsi} and sensing \cite{teng2025sensing,wangdrlfim2026}. Nevertheless, the potential of employing FIM for PLS remains largely unexplored. 

Motivated by this gap, this paper investigates FIM as a transmitter to combat multi-antenna eavesdropping. The main contributions of this paper are summarized as follows:
\begin{itemize}
	\item We propose an FIM-aided secure MISO communication system. An SRM problem is formulated by jointly optimizing the transmit covariance matrix and the FIM surface shape under transmit power and maximum morphing range constraints.
	\item For a given FIM surface shape, we obtain the closed-form rank-one optimal transmit covariance matrix. For a fixed transmit covariance matrix, a projected gradient ascent (PGA) method is developed to optimize the FIM surface shape. Based on these results, a block coordinate ascent (BCA) algorithm is proposed for the joint design.
	\item For the special line-of-sight (LoS) scenario, we derive a globally optimal closed-form solution under a sufficient morphing range condition. 
	\item Numerical results demonstrate that the proposed FIM design effectively improves the secrecy rate compared with conventional RAA and benchmark beamforming schemes.
\end{itemize}
\section{System Model}
\begin{figure}[!t]
	\centering
	\includegraphics[width=0.45\textwidth]{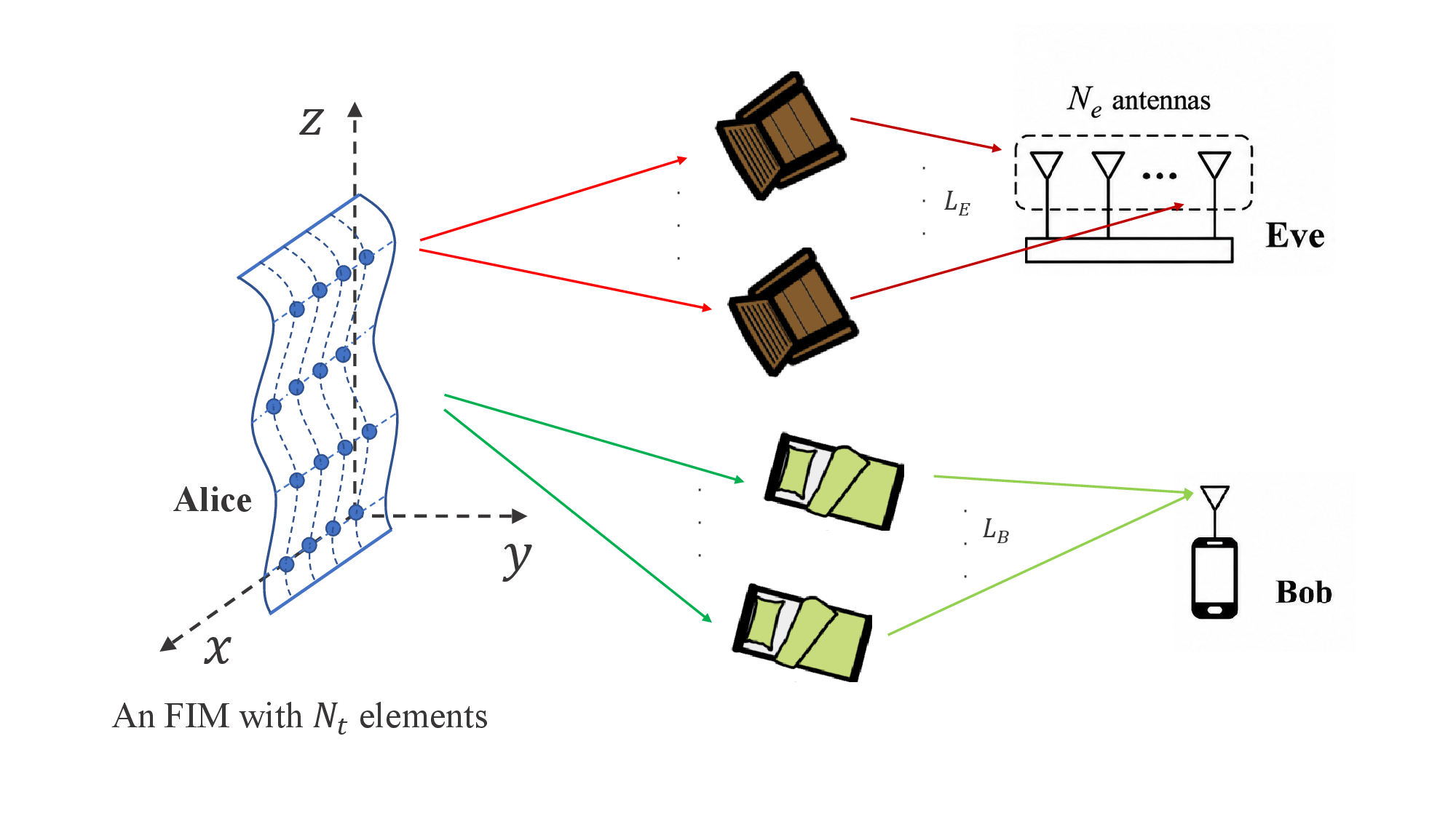}\vspace{-5mm}
	\caption{Illustration of the FIM-assisted secure MISO communication system with a multi-antenna eavesdropper.}\vspace{-5mm}
	\label{fig:system_model}
\end{figure}

As illustrated in Fig. \ref{fig:system_model}, we consider a single-eavesdropper secure communication system, where a base station (BS), referred to as Alice, communicates confidential information with a single-antenna legitimate user, referred to as Bob, in the presence of a multi-antenna eavesdropper, referred to as Eve. Alice is equipped with an FIM comprising $N_t$ radiating elements, while Eve employs $N_e$ receive antennas to model a strong eavesdropping capability via spatial combining. The complex baseband signals in the $t$-th time slot received at Bob and Eve are respectively given by
\begin{align}
	y_b(t) \!=\! \boldsymbol h^H(\boldsymbol d)\boldsymbol x(t)+n(t),
	\quad
	\boldsymbol y_e(t)\! =\! \boldsymbol G(\boldsymbol d)\boldsymbol x(t)+\boldsymbol v(t),
\end{align}
where $\boldsymbol x(t)\in\mathbb{C}^{N_t}$ denotes the transmit signal vector at Alice. Moreover, $n(t)$ and $\boldsymbol v(t)\in\mathbb{C}^{N_e}$ denote the additive white Gaussian noise (AWGN) at Bob and Eve, respectively, satisfying $\mathbb{E}\left\{|n(t)|^2\right\}=\sigma_b^2$, $\mathbb{E}\left\{\boldsymbol v(t)\boldsymbol v^H(t)\right\} =\sigma_e^2\boldsymbol I$, where $\sigma_b^2$ and $\sigma_e^2$ denote the corresponding noise powers.

The channel vector from Alice to Bob is denoted by
$\boldsymbol h(\boldsymbol d)\in\mathbb{C}^{N_t}$, while the channel matrix from Alice to Eve is denoted by
$\boldsymbol G(\boldsymbol d)\in\mathbb{C}^{N_e\times N_t}$.
We assume that the undeformed FIM lies on the $x$-$z$ plane and that each radiating element can morph along the $y$-axis. Let $d_n$ denote the displacement of the $n$th FIM element. In practice, FIM morphing may be constrained by surface continuity, curvature, actuator coupling, and morphing speed. To reveal the fundamental potential of geometric reconfiguration, we adopt an independent element-wise model with $\boldsymbol d=[d_1,d_2,\ldots,d_{N_t}]^T$ and $|d_n|\le d_{\max},\,n=1,\ldots,N_t$, where $d_{\max}$ denotes the maximum morphing range determined by the physical deformation capability of the flexible substrate.

For an azimuth angle $\theta$ and an elevation angle $\varphi$, the undeformed FIM is modeled as a uniform planar array (UPA) on the $x$--$z$ plane. Accordingly, its array response vector is expressed as $\boldsymbol a_{\rm U}(\theta,\varphi)=\boldsymbol a_x(\theta,\varphi)\otimes\boldsymbol a_z(\varphi)\in\mathbb{C}^{N_t}$, where $ \otimes $ denotes the Kronecker product, and
$\boldsymbol a_x(\theta,\varphi)=\left[1,\ldots,e^{-jk_cd_x(N_x-1)\sin\theta\cos\varphi}\right]^T$, $\boldsymbol a_z(\varphi)=\left[1,\ldots,e^{-jk_cd_z(N_z-1)\cos\varphi}\right]^T$. Here, $N_x$ and $N_z$ denote the numbers of FIM elements along the $x$- and $z$-axes, respectively, with $N_t=N_xN_z$. Moreover, $d_x$ and $d_z$ denote the corresponding inter-element spacings, $k_c=2\pi/\lambda$ is the wavenumber, and $\lambda$ is the carrier wavelength.

Since each FIM element can morph along the $y$-axis, the deformation introduces an additional phase response given by $\boldsymbol a_d(\theta,\varphi,\boldsymbol d)=\left[e^{-jk_cd_1\sin\theta\sin\varphi},\ldots, e^{-jk_cd_{N_t}\sin\theta\sin\varphi}\right]^T$. Therefore, the actual array response of the morphed FIM is given by $\boldsymbol a(\theta,\varphi,\boldsymbol d)=\boldsymbol a_{\rm U}(\theta,\varphi) \odot\boldsymbol a_d(\theta,\varphi,\boldsymbol d)$, where $ \odot $ denotes the Hadamard product.

Based on the above FIM array response model, the multipath MISO channel from Alice to Bob is modeled as
\begin{equation}
	\boldsymbol h(\boldsymbol d)=\sum_{\ell_B=1}^{L_B}\beta_{\ell_B}\boldsymbol a(\theta_{B,\ell_B},\varphi_{B,\ell_B},\boldsymbol d)\in\mathbb{C}^{N_t},\label{eq:bob_channel}
\end{equation}
where $L_B$ denotes the number of propagation paths between Alice and Bob, $\beta_{\ell_B}$ is the complex channel gain of the $\ell_B$-th path, and $\theta_{B,\ell_B}$ and $\varphi_{B,\ell_B}$ denote the corresponding azimuth and elevation angles, respectively, $\ell_B\in\{1,2,\dots, L_B\}$.

Since Eve is equipped with $N_e$ receive antennas, the multipath multiple-input multiple-output (MIMO) channel between Alice and Eve is modeled as
\begin{align}
	\boldsymbol G(\boldsymbol d)=\boldsymbol A_E\boldsymbol\Gamma\boldsymbol A^H(\boldsymbol d)\in\mathbb{C}^{N_e\times N_t}.
	\label{eq:eve_channel}
\end{align}
The matrices $\boldsymbol A_E\in\mathbb{C}^{N_e\times L_E}$, $\boldsymbol A(\boldsymbol d)\in
\mathbb{C}^{N_t\times L_E}$ in \eqref{eq:eve_channel} are respectively defined as
\begin{align}
	\boldsymbol A_E&=\left[\boldsymbol a_E\left(\theta_{E,1}^{\rm r},\varphi_{E,1}^{\rm r}\right),\ldots,\boldsymbol a_E\left(\theta_{E,L_E}^{\rm r},\varphi_{E,L_E}^{\rm r}\right)\right],\\
	\boldsymbol A(\boldsymbol d)&=\left[\boldsymbol a\left(\theta_{E,1}^{\rm t},\varphi_{E,1}^{\rm t},\boldsymbol d\right),\!\ldots\!,\boldsymbol a\left(\theta_{E,L_E}^{\rm t},\varphi_{E,L_E}^{\rm t},\boldsymbol d\right)\right],
\end{align}
and $\boldsymbol \Gamma=\operatorname{diag}\left(\gamma_1,\gamma_2,\ldots,\gamma_{L_E}\right)$, where $L_E$ denotes the number of propagation paths between Alice and Eve. $\gamma_{\ell_E}$ is the complex gain of the $\ell_E$-th eavesdropping path. Moreover, $\theta_{E,\ell_E}^{\rm t}$ and $\varphi_{E,\ell_E}^{\rm t}$ denote the departure angles of the $\ell_E$-th path at the FIM, while $\theta_{E,\ell_E}^{\rm r}$ and $\varphi_{E,\ell_E}^{\rm r}$ denote the corresponding arrival angles at Eve, $\ell_E\in\{1,2,\dots, L_E\}$. The receive array response vector of Eve is denoted by $\boldsymbol a_E(\theta,\varphi)\in\mathbb{C}^{N_e}$.

\section{SRM Against Multi-Antenna Eavesdropping}
We assume that Alice has perfect knowledge of the channels $\boldsymbol h(\boldsymbol d)$ and $\boldsymbol G(\boldsymbol d)$. This assumption is adopted to reveal the fundamental secrecy potential of FIM. Let $\boldsymbol W=\mathbb{E}\left\{\boldsymbol x(t)\boldsymbol x^H(t)\right\}\succeq\boldsymbol 0$ denote the transmit covariance matrix, which is subject to the total transmit power constraint $\operatorname{tr}\left(\boldsymbol W\right)\le P$,
where $P$ denotes the maximum transmit power at Alice. For a given transmit covariance matrix $\boldsymbol W$ and FIM shape $\boldsymbol d$, the achievable secrecy rate is given by
\begin{align}
	R_s\left(\boldsymbol W,\boldsymbol d\right)
	&=\left[\log_2\left(1+\frac{\bm h^H(\bm d) \bm W\bm h(\bm d)}{\sigma_b^2}\right)\right.\nonumber\\
	&\left.-\log_2\det\left(\bm I+\frac{\bm G(\bm d) \bm W\bm G^H(\bm d)}{\sigma_e^2}\right)\right]^+.
	\label{eq:secrecy_rate_func}
\end{align}
Accordingly, the SRM problem for the considered single-eavesdropper scenario is formulated as
\begin{subequations}
	\label{eq:srm_single_eve}
		\begin{align}
		\max_{\bm W,\bm d}\quad &R_s\left(\boldsymbol W,\boldsymbol d\right) \\
		\text{s.t.}\quad & \bm W\succeq \bm 0,\label{sd}\\
		 &\mathrm{tr}(\bm W)\leq P,\label{powconstriant}\\
		& |d_n|\le d_{\max},\qquad n=1,\ldots,N_t\label{dconstriant}.
	\end{align}
\end{subequations}
Problem \eqref{eq:srm_single_eve} is non-convex due to the coupled dependence of $\boldsymbol h(\boldsymbol d)$ and $\boldsymbol G(\boldsymbol d)$ on the FIM surface shape, as well as the difference-of-logarithmic terms in the objective function. Consequently, jointly optimizing $\boldsymbol W$ and $\boldsymbol d$ is challenging. In the following, we first characterize the optimal transmit covariance matrix for a given FIM surface shape and then optimize the FIM surface shape for a fixed transmit covariance matrix. Based on the resulting subproblem solutions, a BCA framework is developed for solving \eqref{eq:srm_single_eve}.
\subsection{Optimal Transmit Covariance Matrix Design}

For a given FIM surface shape $\boldsymbol d$, the SRM problem \eqref{eq:srm_single_eve} reduces to the following transmit covariance optimization problem:
\begin{subequations}\label{prob}
\begin{align}
	\max_{\boldsymbol W}\quad&R_s\left(\boldsymbol W,\boldsymbol d\right)\\
	\mathrm{s.t.}\quad&\eqref{sd},\eqref{powconstriant}.
\end{align}
\end{subequations}
Although subproblem \eqref{prob} is still non-convex, the optimal solution to problem \eqref{prob} can be characterized in the following lemma \cite{li2011optimal}:

\emph{Lemma 1}: For a given FIM surface shape $\boldsymbol d$, the optimal transmit covariance matrix $\boldsymbol W^\star$ of problem \eqref{prob} has rank at most one, i.e.,
	\begin{equation}
		\boldsymbol W^\star=
		\begin{cases}
			P\boldsymbol w^\star{\boldsymbol w^\star}^H,&\lambda_{\max}>1\\
			\boldsymbol 0,&\lambda_{\max}\le1
		\end{cases},
	\end{equation}
where $\lambda_{\max}$ is the largest generalized eigenvalue of $\left(\boldsymbol I+\frac{P}{\sigma_b^2}\boldsymbol h(\boldsymbol d)\boldsymbol h^H(\boldsymbol d),\boldsymbol I+\frac{P}{\sigma_e^2}\boldsymbol G^H(\boldsymbol d)\boldsymbol G(\boldsymbol d)\right)$, and $\boldsymbol w^\star$ is the unit-norm generalized eigenvector corresponding to $\lambda_{\max}$.

\subsection{Optimal FIM Surface Shape Design}

For a given transmit covariance matrix $\boldsymbol W$, the SRM problem \eqref{eq:srm_single_eve} reduces to the following FIM surface shape optimization problem:
\begin{align}
	\max_{\boldsymbol d}\quad&R_s(\boldsymbol W,\boldsymbol d)\label{prob:optimal_d}\\
	\mathrm{s.t.}\quad&\eqref{dconstriant}.
\end{align}
Problem \eqref{prob:optimal_d} is non-convex due to the nonlinear dependence of the legitimate and eavesdropping channels on the FIM surface shape. To obtain a high-quality solution, we employ the PGA algorithm to iteratively optimize $\boldsymbol d$.

For notational convenience, define $F_1(\boldsymbol d)=\log_2\left(1+{\boldsymbol h^H(\boldsymbol d)\boldsymbol W\boldsymbol h(\boldsymbol d)}/{\sigma_b^2}\right)$, $F_2(\boldsymbol d)=\log_2\det\left(\boldsymbol I+{\boldsymbol G(\boldsymbol d)\boldsymbol W\boldsymbol G^H(\boldsymbol d)}/{\sigma_e^2}\right)$.
The gradient of the secrecy rate with respect to the FIM shape vector is given by $\nabla_{\boldsymbol d}R_s(\boldsymbol W,\boldsymbol d)=\nabla_{\boldsymbol d}F_1(\boldsymbol d)-\nabla_{\boldsymbol d}F_2(\boldsymbol d)$. The detailed derivation of the gradients is provided in Appendix \ref{app:gradient_derivation}.

Accordingly, at each PGA iteration, the FIM shape vector is updated as
\begin{equation}
	\boldsymbol d\leftarrow\Pi_{\mathcal D}\left(\boldsymbol d+\eta\nabla_{\boldsymbol d}R_s\left(\boldsymbol W,\boldsymbol d\right)\right),
	\label{eq:pga_update}
\end{equation}
where $\eta>0$ denotes the step size and is determined via Armijo backtracking line search. $\Pi_{\mathcal D}(\cdot)$ is the projection onto the feasible set $\mathcal D$. Specifically, for any vector $\boldsymbol z$, the projection is performed element-wise as $\left[\Pi_{\mathcal D}(\boldsymbol z)\right]_n=\min\left\{d_{\max},\max\left\{-d_{\max},z_n\right\}\right\}$.
\subsection{The Proposed BCA Algorithm}
The proposed joint design is solved by a BCA algorithm with two outer optimization blocks. 
For initialization, $G$ feasible FIM surface shapes are randomly generated within the morphing space, and the one yielding the largest secrecy rate, with its corresponding optimal transmit covariance matrix obtained from Lemma~1, is selected as the initial configuration $\boldsymbol d^{(0)}$.
In the $r$-th outer iteration, $\boldsymbol W$ is first updated according to Lemma~1 for the fixed FIM surface shape $\boldsymbol d^{(r)}$. Then, for the obtained $\boldsymbol W^{(r+1)}$, the FIM surface shape is optimized by the inner PGA procedure in \eqref{eq:pga_update}, resulting in $\boldsymbol d^{(r+1)}$. The outer BCA loop is repeated until the secrecy rate improvement between two consecutive outer iterations is below a prescribed threshold $\epsilon$, i.e., $|R_s(\boldsymbol W^{(r+1)},\boldsymbol d^{(r+1)})-R_s(\boldsymbol W^{(r)},\boldsymbol d^{(r)})|\leq\epsilon$. The overall computational complexity of the proposed BCA algorithm is $\mathcal{O}(I_{\rm BCA}[N_t^3+N_eN_t^2+I_{\rm PGA}(N_e^3+N_tN_e(N_t+N_e+L_E)+N_tL_B)])$, where $I_{\rm BCA}$ and $I_{\rm PGA}$ denote the numbers of BCA and PGA iterations, respectively.
\section{Globally Optimal Closed-Form Design for the LoS Scenario}

To reveal the fundamental role of FIM morphing in secure communications, we consider a special pure LoS scenario. From the transmitter, Bob and Eve are observed at the departure angles $(\theta_B,\varphi_B)$ and $(\theta_E^{\rm t},\varphi_E^{\rm t})$, respectively, while the LoS signal impinges on Eve from the arrival angles $(\theta_E^{\rm r},\varphi_E^{\rm r})$. The corresponding channels are given by $\boldsymbol h(\boldsymbol d)=\alpha_b\boldsymbol a(\theta_B,\varphi_B,\boldsymbol d)$, $\boldsymbol G(\boldsymbol d)=\alpha_e\boldsymbol a_E(\theta_E^{\rm r},\varphi_E^{\rm r})\boldsymbol a^H(\theta_E^{\rm t},\varphi_E^{\rm t},\boldsymbol d)$, where $\alpha_b$ and $\alpha_e$ denote the complex LoS path gains. The following proposition provides a globally optimal closed-form solution of the SRM problem under a sufficient morphing-range condition.

\emph{Proposition 1}: In the LoS scenario, define $\Delta_{b,e}=\sin\theta_E^{\rm t}\sin\varphi_E^{\rm t}-\sin\theta_B\sin\varphi_B$. If $N_t\ge2$ and $d_{\max}\ge{\lambda}/{2|\Delta_{b,e}|}$, a globally optimal solution to problem \eqref{eq:srm_single_eve} is the maximum-ratio transmission (MRT)
	\begin{equation}
		\boldsymbol W^\star=P\frac{\boldsymbol h(\boldsymbol d^\star)\boldsymbol h^H(\boldsymbol d^\star)}{\|\boldsymbol h(\boldsymbol d^\star)\|^2},
		\label{eq:LoS_optimal_W}
	\end{equation}
	where 
	\begin{align}
		d_n^\star=\frac{\psi_n-\delta_n+2\pi k_n}{k_c\Delta_{b,e}}, n=1,\ldots,N_t,\label{closed}
	\end{align}
	with $\psi_n=\psi_0+{2\pi(n-1)}/{N_t}$, where $\psi_0$ is an arbitrary common phase and $k_n\in\mathbb Z$ is selected such that $|d_n^\star|\le d_{\max}$. Moreover, $\delta_n=k_c[x_n(\sin\theta_E^{\rm t}\cos\varphi_E^{\rm t}-\sin\theta_B\cos\varphi_B)+z_n(\cos\varphi_E^{\rm t}-\cos\varphi_B)]$,
	where $x_n=d_x\left\lfloor(n-1)/N_z\right\rfloor$, $z_n=d_z\operatorname{mod}(n-1,N_z)$, and the globally optimal secrecy rate is $R_s^\star=\log_2\left(1+{PN_t|\alpha_b|^2}/{\sigma_b^2}\right)$.

\emph{Proof:} See Appendix~\ref{app:proof_los_theorem}.

Proposition 1 also shows that a larger morphing range is required as Bob and Eve become angularly closer. When Bob and Eve are aligned in the same transmit direction, $\Delta_{b,e}=0$, making PLS infeasible.

\section{Simulation Results}
In this section, numerical results are presented to evaluate the effectiveness of the proposed FIM-aided secure transmission scheme.
The undeformed FIM is configured as a UPA on the $x$--$z$ plane, with half-wavelength element spacing along both the $x$- and $z$-axes. Unless otherwise specified, Bob is equipped with a single antenna, while Eve employs a UPA with half-wavelength element spacing. Each FIM element can independently morph along the $y$-axis, with $d_{\max}=0.5\lambda$. The noise powers at Bob and Eve are set to $\sigma_b^2=\sigma_e^2=-94$ dBm. The large-scale channel power gains are modeled as $\chi_B^2=\chi_0^2(r_B/r_0)^{-2.2}$ and $\chi_E^2=\chi_0^2(r_E/r_0)^{-2.2}$, where $\chi_0^2=-40$ dB denotes the reference path loss at $r_0=1$ m. The distances from Alice to Bob and Eve are set to $r_B=r_E=100$ m. $\beta_{\ell_B}\sim\mathcal{CN}(0,\chi_B^2/L_B)$, $\gamma_{\ell_E}\sim\mathcal{CN}(0,\chi_E^2/L_E)$, where $\mathcal{CN}(\mu,\sigma^2)$ represents the complex Gaussian distribution with mean $\mu$ and variance $\sigma^2$. $\theta_{B,\ell_B},\varphi_{B,\ell_B},\theta_{E,\ell_E}^{\rm t},\varphi_{E,\ell_E}^{\rm t},\theta_{E,\ell_E}^{\rm r},\varphi_{E,\ell_E}^{\rm r}\sim\mathcal{U}(0,\pi)$, where $\mathcal{U}(a,b)$ represents the uniform distribution over the interval $(a,b)$. We set $N_t=16$, $N_e=9$, and $L_B=L_E=5$. The system operates at 28 GHz. All simulation results are averaged over 100 independent channel realizations.

For comparison, we consider both the proposed FIM and a conventional rigid antenna array (RAA), where the RAA is modeled as a fixed UPA with half-wavelength element spacing. Three transmit beamforming strategies are considered, namely MRT, zero-forcing (ZF), and the optimal transmit design. Accordingly, the benchmark schemes include RAA-MRT, RAA-ZF, RAA-OPT, FIM-MRT, FIM-ZF, and FIM-OPT. In particular, OPT denotes the optimal transmit covariance design obtained from the dominant generalized eigenvector.
\subsection{Secrecy Rate Versus Transmit Power}

Fig. \ref{pow} shows the secrecy rate versus the transmit power for different array architectures and beamforming schemes. It is observed that the secrecy rate increases with the transmit power for all schemes. MRT suffers from severe information leakage because it only maximizes the received signal power at Bob without explicitly suppressing Eve. ZF can reduce the leakage toward Eve, but this comes at the cost of a reduced legitimate beamforming gain. OPT provides a better balance between Bob signal enhancement and Eve leakage suppression, and therefore outperforms both MRT and ZF. Compared with RAA-OPT, FIM-OPT achieves a transmit-power saving of approximately $2.5$ dB at the same secrecy rate.

\subsection{Secrecy Rate Versus Maximum Morphing Range}
\begin{figure}[]
	\centering
	\subfigure[]{
		\label{pow}
		\includegraphics[width=0.244\textwidth]{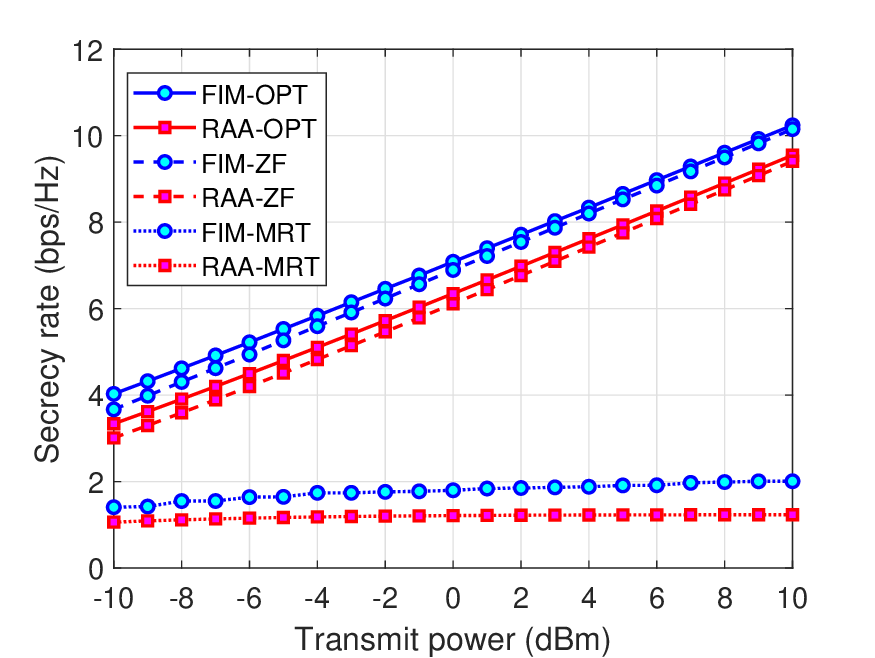}
	}\hspace{-0.6cm}
	\subfigure[]{
		\label{range}
		\includegraphics[width=0.244\textwidth]{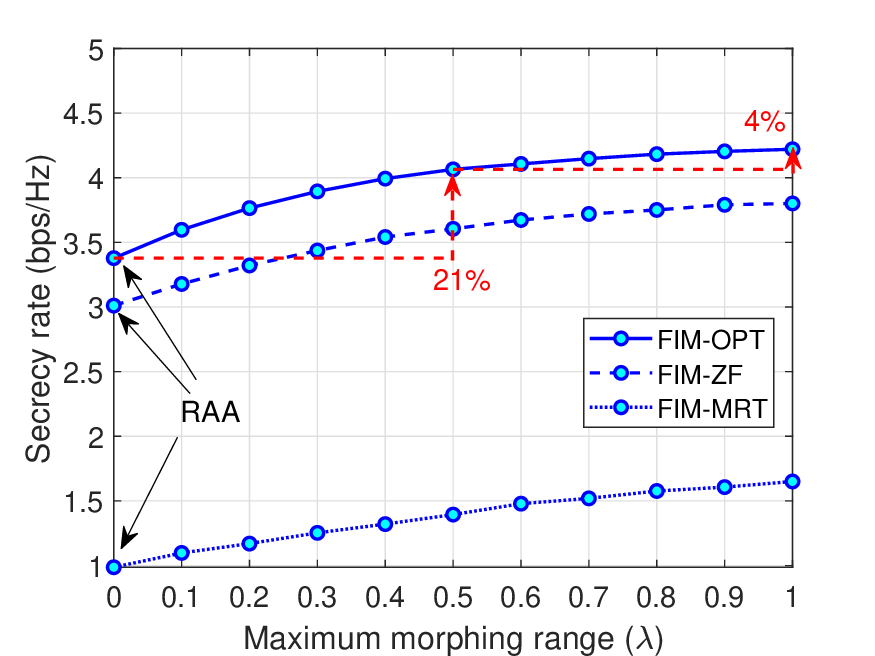}
	}\vspace{-0.25cm}
	\caption{(a) Secrecy rate versus transmit power under different benchmark schemes;
		(b) Secrecy rate versus maximum morphing range under different benchmark schemes.}\vspace{-0.55cm}
\end{figure}

Fig. \ref{range} illustrates the secrecy rate versus the maximum morphing range $d_{\max}$ when $P=-10$ dBm. When $d_{\max}=0$, the FIM degenerates into a conventional RAA. As $d_{\max}$ increases, the secrecy rate of the FIM-based schemes gradually improves, which demonstrates the benefit of introducing geometry-domain degrees of freedom. It can also be observed that the secrecy-rate improvement becomes saturated when $d_{\max}$ is sufficiently large. Specifically, in the FIM-OPT scheme, increasing $d_{\max}$ from $0$ to $0.5\lambda$ brings about a $21\%$ secrecy-rate improvement, while further increasing $d_{\max}$ from $0.5\lambda$ to $\lambda$ only provides an additional $4\%$ gain. This is because once the FIM has obtained enough morphing capability to create a favorable channel separation between Bob and Eve, further increasing the morphing range provides only limited performance gains. Thus, a moderate morphing range can already capture most of the FIM secrecy gain in practice.

\subsection{Verification of the Closed-Form Design in the LoS Scenario}

\begin{figure*}[]
	\centering
	\subfigure[]{
		\label{RAA_MRT}
		\includegraphics[width=0.31\textwidth]{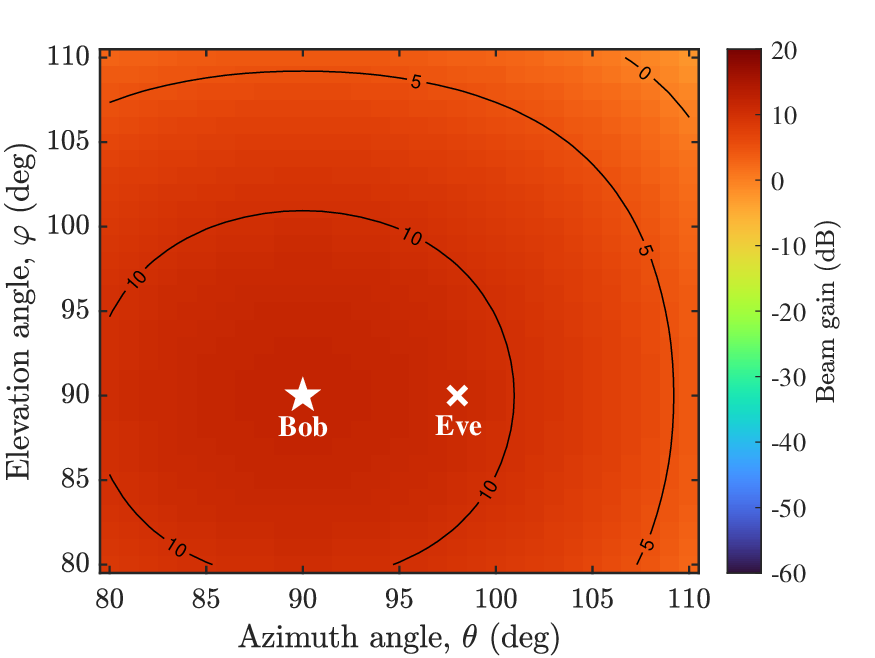}
	}
	\subfigure[]{
		\label{RAA_ZF}
		\includegraphics[width=0.31\textwidth]{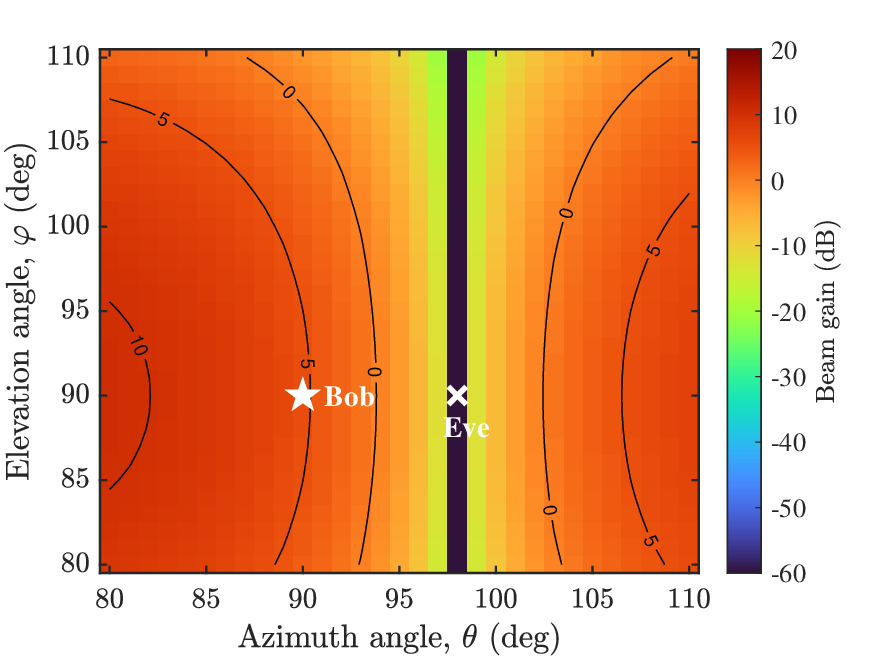}
	}
	\subfigure[]{
		\label{FIM_MRT}
		\includegraphics[width=0.31\textwidth]{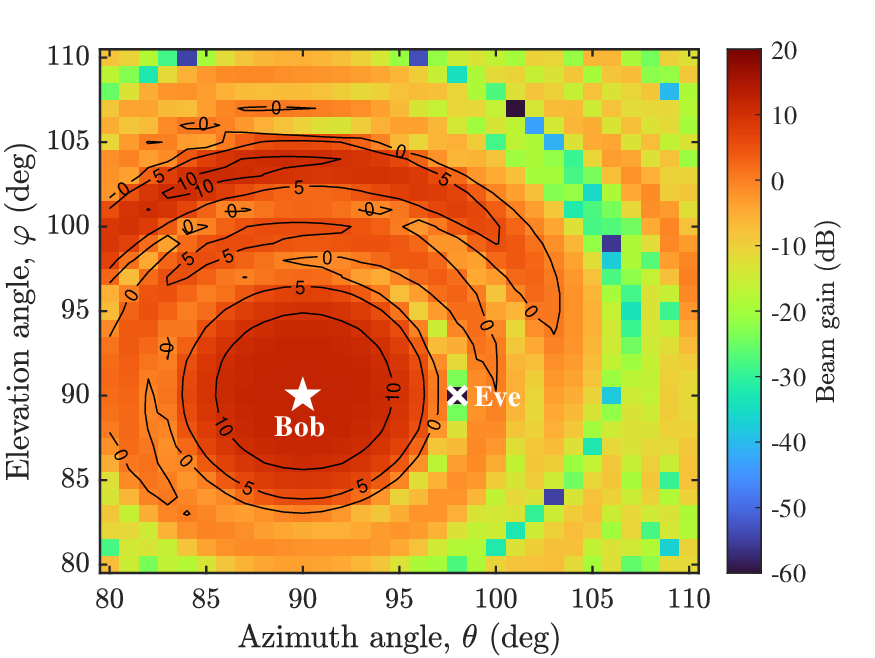}
	}\vspace{-0.25cm}
	\caption{Beampattern comparison under different benchmark schemes,
		where (a) RAA-MRT;
		(b) RAA-ZF;
		(c) FIM-MRT (closed-form).}\vspace{-0.55cm}
	\label{bp}
\end{figure*}

Fig. \ref{bp} verifies the closed-form design derived in Proposition 1 under the LoS single-antenna Eve scenario. In this example, the single-antenna Bob and Eve are located at $(\theta_B,\varphi_B)=(90^\circ,90^\circ)$ and $(\theta_E^{\rm t},\varphi_E^{\rm t})=(98^\circ,90^\circ)$, respectively. Since Bob and Eve are closely located in the angular domain, their array responses are highly correlated for a conventional RAA, which makes secure beamforming challenging. It is observed that RAA-MRT generates high beam gains toward both Bob and Eve. This is because MRT only maximizes the legitimate received signal power without explicitly suppressing the information leakage toward Eve. As a result, although RAA-MRT achieves a large array gain at Bob, it also causes severe signal leakage in the direction of Eve. In contrast, RAA-ZF successfully forms a null at Eve, thereby eliminating the eavesdropping signal. However, this nulling operation is achieved at the cost of more than $6$ dB array-gain loss. By comparison, the closed-form FIM-MRT design forms a deep null toward Eve while maintaining the maximum beam gain at Bob. 

\section{Conclusion}

In this paper, we investigated FIM-aided secure MISO communications in the presence of a multi-antenna eavesdropper. By exploiting the geometry reconfigurability of the FIM, the legitimate and eavesdropping channels can be proactively reshaped to improve PLS. We formulated an SRM problem by jointly optimizing the transmit covariance matrix and the FIM surface shape. For a given FIM shape, the optimal transmit covariance matrix was characterized as a rank-one solution obtained from the dominant generalized eigenvector. For a fixed transmit covariance matrix, a PGA method was developed to optimize the FIM surface shape, leading to a BCA algorithm for the joint design. Furthermore, for the special LoS scenario, we derived a globally optimal closed-form solution under a sufficient morphing range condition. Numerical results demonstrated that the proposed scheme effectively outperforms conventional RAA-based beamforming schemes. 

\begin{appendices}
	\section{Gradient Derivation for FIM Surface Shape Optimization}
	\label{app:gradient_derivation}
	
	We first derive the gradient of the eavesdropping rate $F_2(\boldsymbol d)$. Define $\boldsymbol Q_e=\boldsymbol I+{\boldsymbol G(\boldsymbol d)\boldsymbol W\boldsymbol G^H(\boldsymbol d)}/{\sigma_e^2}\in\mathbb{C}^{N_e\times N_e}$. The partial derivative of $F_2(\boldsymbol d)$ with respect to $d_n$ is given by \cite{hjorungnes2007complex}
	\begin{equation}
		\frac{\partial F_2}{\partial d_n}=\frac{1}{\ln2}\operatorname{tr}\left(\boldsymbol Q_e^{-1}\frac{\partial\boldsymbol Q_e}{\partial d_n}\right),
		\label{eq:appendix_partial_F2}
	\end{equation}
	where ${\partial\boldsymbol Q_e}/{\partial d_n}\in\mathbb{C}^{N_e\times N_e}$ is given by
	\begin{equation}
		\frac{\partial\boldsymbol Q_e}{\partial d_n}\!=\!\frac{1}{\sigma_e^2}\left(\frac{\partial\boldsymbol G(\boldsymbol d)}{\partial d_n}\boldsymbol W\boldsymbol G^H(\boldsymbol d)+\boldsymbol G(\boldsymbol d)\boldsymbol W\frac{\partial\boldsymbol G^H(\boldsymbol d)}{\partial d_n}\right).
		\label{eq:appendix_partial_Qe}
	\end{equation}
	Recall that $\boldsymbol G(\boldsymbol d)=\boldsymbol A_E\boldsymbol\Gamma\boldsymbol A^H(\boldsymbol d)$. Since $\boldsymbol A_E$ and $\boldsymbol\Gamma$ are independent of $\boldsymbol d$, ${\partial\boldsymbol G(\boldsymbol d)}/{\partial d_n}\in\mathbb{C}^{N_e\times N_t}$ and ${\partial\boldsymbol G^H(\boldsymbol d)}/{\partial d_n}\in\mathbb{C}^{N_t\times N_e}$ can be expressed as
	\begin{align}
		\frac{\partial\boldsymbol G(\boldsymbol d)}{\partial d_n}=\boldsymbol A_E\boldsymbol\Gamma\frac{\partial\boldsymbol A^H(\boldsymbol d)}{\partial d_n},\quad\frac{\partial\boldsymbol G^H(\boldsymbol d)}{\partial d_n}=\frac{\partial\boldsymbol A(\boldsymbol d)}{\partial d_n}\boldsymbol\Gamma^H\boldsymbol A^H_E.\label{eq:appendix_partial_G}
	\end{align}
	According to the FIM array response model, we have
	\begin{equation}
		\frac{\partial\boldsymbol A(\boldsymbol d)}{\partial d_n}=-jk_c\widetilde{\boldsymbol A}_n\operatorname{diag}\left(\boldsymbol s_{\theta}\odot\boldsymbol s_{\varphi}\right)\in\mathbb{C}^{N_t\times L_E},
		\label{eq:appendix_partial_A}
	\end{equation}
	where $\boldsymbol s_{\theta}=\left[\sin\theta_{E,1}^{\rm t},\ldots,\sin\theta_{E,L_E}^{\rm t}\right]^T\in\mathbb{C}^{L_E}$, $\boldsymbol s_{\varphi}=\left[\sin\varphi_{E,1}^{\rm t},\ldots,\sin\varphi_{E,L_E}^{\rm t}\right]^T\in\mathbb{C}^{L_E}$,
	and $\widetilde{\boldsymbol A}_n=[\boldsymbol 0_{(n-1)\times L_E};\boldsymbol A_{n,:}(\boldsymbol d);\boldsymbol 0_{(N_t-n)\times L_E}]$, $\boldsymbol A_{n,:}(\boldsymbol d)$ denotes the $n$-th row of $\boldsymbol A(\boldsymbol d)$.
	Substituting \eqref{eq:appendix_partial_A} into \eqref{eq:appendix_partial_G}, \eqref{eq:appendix_partial_Qe} and subsequently into \eqref{eq:appendix_partial_F2} yields $\partial F_2/\partial d_n$.
	Finally, $\nabla_{\boldsymbol d}F_2(\boldsymbol d)=\left[{\partial F_2}/{\partial d_1},\ldots,{\partial F_2}/{\partial d_{N_t}}\right]^T$. The gradient $\nabla_{\boldsymbol d}F_1(\boldsymbol d)$ can be obtained in the same manner by viewing the legitimate MISO link as a single receive antenna special case, and is omitted for brevity. Therefore, $\nabla_{\boldsymbol d}R_s(\boldsymbol W,\boldsymbol d)=\nabla_{\boldsymbol d}F_1(\boldsymbol d)-\nabla_{\boldsymbol d}F_2(\boldsymbol d)$.

\section{Proof of Proposition 1}
\label{app:proof_los_theorem}

For any feasible $\boldsymbol W$ and $\boldsymbol d$, the non-negativity of the eavesdropping rate gives $R_s(\boldsymbol W,\boldsymbol d)\le\log_2\left(1+{\boldsymbol h^H(\boldsymbol d)\boldsymbol W\boldsymbol h(\boldsymbol d)}/{\sigma_b^2}\right)$.
Since $\boldsymbol W\succeq\boldsymbol 0$ and $\operatorname{tr}(\boldsymbol W)\le P$, $\boldsymbol h(\boldsymbol d)^H\boldsymbol W\boldsymbol h(\boldsymbol d)\le\lambda_{\max}(\boldsymbol W)\|\boldsymbol h(\boldsymbol d)\|^2\le\operatorname{tr}(\boldsymbol W)\|\boldsymbol h(\boldsymbol d)\|^2\le P\|\boldsymbol h(\boldsymbol d)\|^2$. This upper bound is achieved by the MRT covariance matrix $\boldsymbol W=P\boldsymbol h(\boldsymbol d)\boldsymbol h^H(\boldsymbol d)/\|\boldsymbol h(\boldsymbol d)\|^2$.
Under the pure LoS model, $\|\boldsymbol h(\boldsymbol d)\|^2=N_t|\alpha_b|^2$. Therefore, $R_s(\boldsymbol W,\boldsymbol d)\le	\log_2\left(1+{PN_t|\alpha_b|^2}/{\sigma_b^2}\right)$.

Next, the cross-correlation between the array responses of Bob and Eve can be written as $\boldsymbol a^H(\theta_E^{\rm t},\varphi_E^{\rm t},\boldsymbol d)\boldsymbol a(\theta_B,\varphi_B,\boldsymbol d)=\sum_{n=1}^{N_t}e^{j(\delta_n+k_c\Delta_{b,e}d_n)}$. Since $d_n\in[-d_{\max},d_{\max}]$, the tunable phase interval of $k_c\Delta_{b,e}d_n$ has length $2k_c|\Delta_{b,e}|d_{\max}$. When $d_{\max}\ge{\lambda}/{2|\Delta_{b,e}|}$, we have $2k_c|\Delta_{b,e}|d_{\max}={4\pi}|\Delta_{b,e}|d_{\max}/{\lambda}\ge2\pi$.
Hence, any target phase modulo $2\pi$ is achievable for each FIM element. By selecting $d_n^\star$ according to \eqref{closed}, we have $\delta_n+k_c\Delta_{b,e}d_n^\star\equiv\psi_n\pmod{2\pi}$.
It follows that
\begin{align}
	&\boldsymbol a^H(\theta_E^{\rm t},\varphi_E^{\rm t},\boldsymbol d^\star)
	\boldsymbol a(\theta_B,\varphi_B,\boldsymbol d^\star)\nonumber\\
	&=e^{j\psi_0}\sum_{n=1}^{N_t}\exp\left(j\frac{2\pi(n-1)}{N_t}\right)=0\label{eq:LoS_zero_correlation}.
\end{align}
Meanwhile, by applying the MRT beamformer $\boldsymbol W^\star=P\boldsymbol h(\boldsymbol d^\star)\boldsymbol h^H(\boldsymbol d^\star)/\|\boldsymbol h(\boldsymbol d^\star)\|^2$, \eqref{eq:LoS_zero_correlation} implies $\boldsymbol G(\boldsymbol d^\star)\boldsymbol W^\star\boldsymbol G^H(\boldsymbol d^\star)=\boldsymbol0$.
Therefore, $R_s(\boldsymbol W^\star,\boldsymbol d^\star)=\log_2\left(1+{PN_t|\alpha_b|^2}/{\sigma_b^2}\right)$,
which completes the proof.
\hfill$\blacksquare$
\end{appendices}
\bibliographystyle{IEEEtran}
\bibliography{Reference}

@article{an2023stacked,
  title={Stacked intelligent metasurface-aided {MIMO} transceiver design},
  author={An, Jiancheng and Yuen, Chau and Xu, Chao and Li, Hongbin and Ng, Derrick Wing Kwan and Di Renzo, Marco and Debbah, Mérouane and Hanzo, Lajos},
  journal={IEEE Wireless Commun.}, 
  year={2024},
  volume={31},
  number={4},  
  month={Aug.},
  pages={123-131}}

@article{hjorungnes2007complex,
  title={Complex-valued matrix differentiation: Techniques and key results},
  author={Hjorungnes, A. and Gesbert, D.},
  journal={IEEE Trans. Signal Process.},
  volume={55},
  number={6},
  pages={2740--2746},
  month={Jun.},
  year={2007}
}

@ARTICLE{anfim2025,
  author={An, Jiancheng and Yuen, Chau and Renzo, Marco Di and Debbah, Mérouane and Vincent Poor, H. and Hanzo, Lajos},
  journal={IEEE Trans. Wireless Commun.}, 
  title={Flexible Intelligent Metasurfaces for Downlink Multiuser {MISO} Communications}, 
  year={2025},
volume={24},
  number={4},
month = {Apr.},
  pages={2940-2955}}

@ARTICLE{teng2025sensing,
  author={Teng, Zihao and An, Jiancheng and Gan, Lu and Al-Dhahir, Naofal and Han, Zhu},
  journal={IEEE Trans. Veh. Technol.}, 
  title={Flexible Intelligent Metasurface for Enhancing Multi-Target Wireless Sensing}, 
  year={2025},
  volume={74},
  number={12},
Month = {Dec.},
  pages={19825-19830}}

@ARTICLE{an2025emer,
  author={An, Jiancheng and Debbah, Mérouane and Cui, Tie Jun and Chen, Zhi Ning and Yuen, Chau},
  journal={IEEE Trans. Antennas Propag.}, 
 title={Emerging Technologies in Intelligent Metasurfaces: Shaping the Future of Wireless Communications}, 
  year={2026},
  volume={74},
  number={5},
  month={May},
  pages={3913-3928}}

@ARTICLE{niu2026survey,
  author={Niu, Hong and Xiao, Yue and Lei, Xia and Chen, Jiangong and Xiao, Zhihan and Li, Mao and Yuen, Chau},
  journal={IEEE Commun. Surveys Tuts.}, 
  title={A Survey on Artificial Noise for Physical Layer Security: Opportunities, Technologies, Guidelines, Advances, and Trends}, 
Month = {Sep.},
  year={2025},
  volume={28},
  pages={341-381}}

@ARTICLE{zhang2025rissecure,
  author={Zhang, Yang and Ren, Hong and Pan, Cunhua and Wang, Boshi and Yu, Zhiyuan and Weng, Ruisong and Wu, Tuo and He, Yongchao},
    journal={IEEE Trans. Veh. Technol.}, 
  title={Secure Wireless Communication in Active {RIS}-Assisted {DFRC} Systems}, 
  year={2025},
  volume={74},
  number={1},
Month = {Jan.},
  pages={626-640}}

@article{li2011optimal,
  author  = {Qiang Li and Wing-Kin Ma},
  title   = {Optimal and Robust Transmit Designs for {MISO} Channel Secrecy by Semidefinite Programming},
  journal = {IEEE Trans. Signal Process.},
  volume  = {59},
  number  = {8},
  pages   = {3799--3812},
  month   = {Aug.},
  year    = {2011},
  doi     = {10.1109/TSP.2011.2146770}
}

@article{niu2025efficient,
  author  = {Hong Niu and Xia Lei and Jiancheng An and Lechen Zhang and Chau Yuen},
  title   = {On the Efficient Design of Stacked Intelligent Metasurfaces for Secure {SISO} Transmission},
  journal = {IEEE Trans. Inf. Forensics and Security},
  volume  = {20},
  pages   = {60--70},
  month   = {Nov.},
  year    = {2024},
}

@article{kavianinia2025secrecy,
  author  = {Mohammad Reza Kavianinia and Abbas Mohammadi and Vahid Meghdadi},
  title   = {Secrecy Rate Maximization in the Presence of Stacked Intelligent Metasurface},
  journal = {IEEE Trans. Inf. Forensics and Security},
  volume  = {20},
  pages   = {6840--6853},
  month   = {Jun.},
  year    = {2025},
  doi     = {10.1109/TIFS.2025.3583242}
}

@article{kumar2026fimstatcsi,
  author  = {V. Kumar and A. Papazafeiropoulos and P. Kourtessis and J. Senior and M. Chafii and D. I. Kaklamani and I. S. Venieris},
  title   = {Flexible Intelligent Metasurface for Downlink Communications Under Statistical {CSI}},
  journal = {IEEE Wireless Commun. Lett.},
  volume  = {15},
  month   = {Dec.},
  pages   = {1150--1154},
  year    = {2025}}

@ARTICLE{wangdrlfim2026,
  author={Wang, Maoyuan and Zhang, Qian and An, Jiancheng and Cheng, Xuejun and Dong, Zheng and Wang, Deqiang},
  journal={IEEE Wireless Commun. Lett.}, 
  title={{DRL}-Based Joint Beamforming and Surface Shape Optimization for Flexible Intelligent Metasurface-Aided {ISAC} Systems}, 
  year={2026},
 note={{E}arly {A}ccess}}
\end{document}